# An Indicator of Research Front Activity: Measuring Intellectual Organization as Uncertainty Reduction in Document Sets

Diana Lucio-Arias [i] and Loet Leydesdorff [ii]

Amsterdam School of Communication Research (ASCoR), University of Amsterdam

Kloveniersburgwal 48, 1012 CX Amsterdam, The Netherlands

**Abstract**

When using scientific literature to model scholarly discourse, a research specialty can be operationalized as an evolving set of related documents. Each publication can be expected to contribute to the further development of the specialty at the research front. The specific combinations of title words and cited references in a paper can then be considered as a signature of the knowledge claim in the paper: new words and combinations of words can be expected to represent variation, while each paper is at the same time selectively positioned into the intellectual organization of a field using context-relevant references. Can the mutual information among these three dimensions—title words, cited references, and sequence numbers—be used as an indicator of the extent to which intellectual organization structures the uncertainty prevailing at a research front? The effect of the discovery of nanotubes (1991) on the previously existing field of fullerenes is used as a test case. Thereafter, this method is applied to science studies with a focus on scientometrics using various sample delineations. An emerging research front about citation analysis can be indicated.

[i] D.P.LucioArias@uva.nl
[ii] loet@leydesdorff.net

# 1. Introduction

The literary footprint of scientific communication in scholarly papers provides us with a map of the process of organizing and controlling the production of scientific knowledge (Price, 1965; Whitley, 1984). Small & Griffith (1974) were the first to operationalize research specialties as cognitive categories indicated by the citation relationships among documents. Garfield *et al*. (1964) proposed to use citations for the historical reconstruction of scientific developments (cf. Pudovkin & Garfield, 2002). In a previous study, we added the notion of evolutionary development in terms of variation and selection of cited references to the algorithmic historiography (Lucio-Arias & Leydesdorff, 2008).

In addition to cited references, scientific texts carry information indicated by selections of words and co-occurrences of words (Callon *et al*., 1986; Leydesdorff, 1991). Words, however, may have different meanings in various contexts (Hesse, 1980; Law & Lodge, 1984; Law, 1986; Leydesdorff, 1997). Whereas cited references provide texts with *contextual* information, title words are carefully selected by authors in order to position their knowledge claims at specific moments of time. Cited references position the text within a socio-cognitive domain along the time dimension, and can thus be expected to operate as codifiers. Words provide variation ("newness") in the discourse, and therefore one can expect words to be less codified than cited references (Leydesdorff, 1989). However, there is no one-to-one relation between title words as variation and cited references as providing the structural and historical contexts; both cited references and title words can vary and be recombined.

New combinations can be expected in terms of both title words and cited references, and in the interactions between these two dimensions. The historical progression generates variation, but the aggregated system can be expected to develop by reorganizing its substantive content continuously. In other words, the history of the system is reflexively rewritten by the discourse (that is, an exchange of arguments and expectations). Citations can be used by authors to reconstruct the history of a field in texts from the perspective of hindsight (Wouters, 1999). As new documents appear at the research front, the past is partially overwritten and forgotten at the supra-individual level of the textual processing of information, meaning, and cognitive content (Kuhn, 1962; Garfield, 1975; Hellsten *et al*., 2006).

Two dynamics along the time axis are thus involved: the historical development of the socio-cognitive system with the arrow of time, and the evolutionary rewriting from the perspective of hindsight which is induced by the further development of and reconstruction by discursive knowledge. In other words, the textual process represents both historical development and its evolutionary restructuring in terms of intellectual organization. A socio-cognitive system of knowledge production and control can be expected to reorganize its content continuously along an emerging axis of codification. This emerging dynamics feeds *back* on the historical development. The feedback potentially reduces uncertainty which is generated with the arrow of time. Can this reduction of uncertainty be measured?

While the development with the axis of time generates probabilistic entropy—because of the Second Law which holds equally for probabilistic entropy[1]—the reconstruction from the perspective of hindsight, that is, against the arrow of time, can be expected to reduce uncertainty in some configurations more than in others. This reduction of uncertainty within a system can be considered as a consequence of the increasing self-organization of discursive knowledge (embedded in texts) at an active research front. Processes of validation in that case organize new contributions intellectually into bodies of knowledge (Lucio-Arias & Leydesdorff, 2009). If the self-organization stagnates, an erosion of structure and therefore an increase of entropy would be expected.

The development of research on fullerenes and nanotubes in the field of nanoscience and nanotechnology provides us with a test case to examine both stability and change at a research front. Fullerenes were discovered in 1985 (Kroto *et al.,* 1985);[2] a Nobel Prize was awarded to Robert F. Curl, Harold W. Kroto, and Richard E. Smalley for this discovery in 1996. Nanotubes were discovered as a specific form of fullerenes in 1991 by Sumio Iijima (Iijima, 1991). After the discovery of nanotubes, the number of publications at this research front developed rapidly, while the number of publications in fullerenes stabilized during the 1990s (Lucio-Arias & Leydesdorff, 2007, at p. 609). Nanotubes are expected to have more technological relevance than fullerenes.

---

[1] Since $S = k_B H$ and $k_B$ is a constant (the Boltzmann constant), the development of $S$ over time is a function of the development of $H$ over time, and *vice versa.*
[2] The first document using fullerenes as one of its title words was published in 1987.

In a previous study, we used HistCite™ for an algorithmic historiography of these two research fronts and showed that the history of fullerenes is more continuous, while the field of nanotubes is still being actively reconstructed from the side of the research front (Lucio-Arias & Leydesdorff, 2008). This study, however, was based on using cited references only. In this study, we extend the reconstruction with the dimension of title words and we use document sets instead of individual documents.

The resulting matrices of title words versus cited references can be expected to contain not only variation but also structure (Braam, 1991; Heimeriks *et al*., 2000; Van den Besselaar & Heimeriks, 2006). Note that these matrices may be sparse because the combinations of title words and cited references in scientific texts are specific. Because of this specificity, one can expect the matrices to be useful as indicators of intellectual organization. Each text is represented as a matrix of title words versus cited references; matrices thereafter are aggregated in yearly sets along the time axis so that one obtains a cube of information as follows:

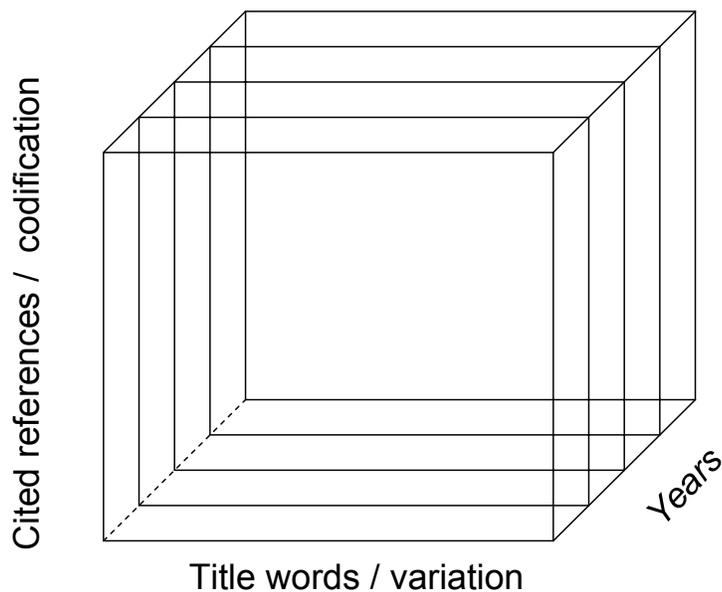

**Figure 1:** Three dimensions considered for the computation of the configurational information.

The three-dimensional array enables us to compute mutual information among three dimensions or, in other words, "configurational information" (McGill, 1954; Abramson, 1963; Jakulin, 2005; Jakulin & Bratko, 2004; Leydesdorff, 2008; Leydesdorff & Sun, 2009; Yeung, 2008). Unlike mutual information between two dimensions, this information measure can be either positive or negative, and thus indicate either an increase or a reduction of the uncertainty prevalent in the set(s) under study.[3] One

---

[3] Both Yeung (2008, p. 59f.) and Krippendorff (2009, p. 200) noted that this information measure can no longer be considered as a Shannon-type measure because of the possible circularity in the information transfers. Shannon-type entropy measures are by definition linear and positive (Leydesdorff, 2009). Since the measure sums Shannon-type measures in terms of bits of information, its dimensionality also consists of bits of information, and therefore it can be used as a measure of uncertainty and uncertainty reduction, respectively (Jakulin & Bratko, 2004; Jakulin, 2005). Yeung

expects new publications to increase uncertainty at the specialty level, but a reduction of uncertainty would indicate the presence of intellectual organization or, in other words, the operation (over time) of an active research front. A research front can be expected to reorganize its past continuously into current socio-cognitive terms and references. Can one thus envisage an indicator for the measurement of cognitive development at the level of a research specialty?

**2. Mutual and configurational information**

Mutual information or transmission $T$ between two dimensions $x$ and $y$ is defined as the difference between the sum of uncertainties in the two probability distributions minus their combined uncertainty, as follows:

$$T_{xy} = H_x + H_y - H_{xy} \qquad (1)$$

in which formula $H_x = -\sum_x p_x \log_2 p_x$ and $H_{xy} = -\sum_{xy} p_{xy} \log_2 p_{xy}$ (Shannon, 1948). When the distributions $\Sigma_x p_x$ and $\Sigma_y p_y$ are independent, $T_{xy} = 0$ and $H_{xy} = H_x + H_y$. In all other cases, $H_{xy} < H_x + H_y$, and therefore $T_{xy}$ is positive (Theil, 1972). The uncertainty which prevails when two probability distributions are combined is reduced by the transmission or mutual information between these distributions.

---

(2008, at pp. 51 ff.) further formalized the configurational information in three or more dimensions into the information measure $\mu^*$.

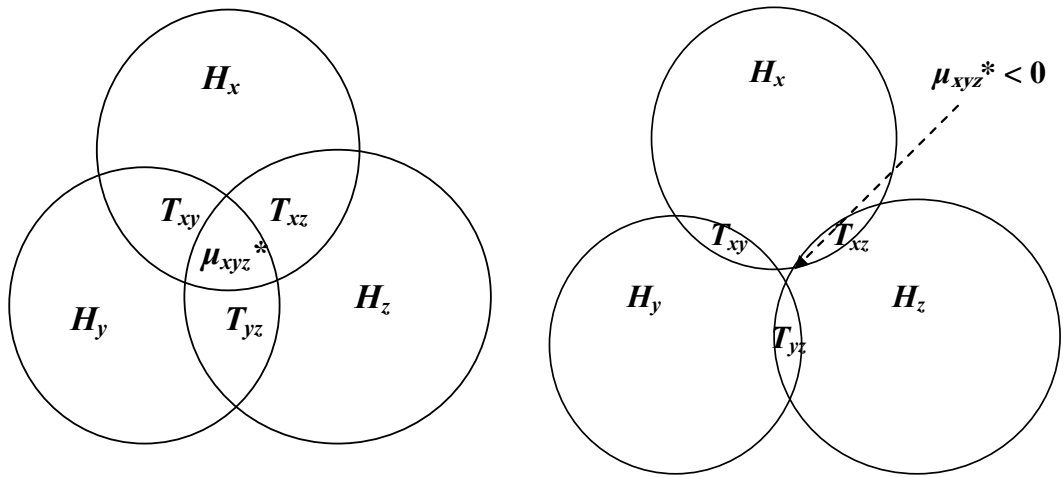

**Figure 2:** Relations between probabilistic entropies (*H*), transmissions (*T*), and configurational information (*μ\**) for three interacting variables.

When three probability distributions are combined, the resulting uncertainty can be positive, zero, or negative depending on the configuration resulting from the interaction terms among these sources. Figure 2 shows the relationships between the probabilistic entropies in each of the dimensions, their mutual information, and the possible configurations. The negative overlap in the right-hand picture ($\mu^* < 0$) illustrates the possibility that the configurational information can disappear or even become negative.

In the left-hand picture, relations are redundant since the same information ($\mu^* > 0$) is received in each dimension from two different sources (Jakulin, 2005). Reduction of uncertainty is generated in the right-hand picture because in addition to the mutual information between *x* and *y*, information can also be transmitted via the third system *z*. This third system can also be considered as part of a next-order loop which feeds back or forward on the transmission between *x* and *y* (Leydesdorff, 2009). If uncertainty is reduced in this feedback system, a synergy is indicated.

McGill (1954) proposed calling this information—albeit with the opposite sign—configurational information. Yeung (2008) specified a corresponding information measure $\mu^*$ which was formalized by Abramson (1963, at p. 129) as follows:

$$\mu^*_{xyz} = H_x + H_y + H_z - H_{xy} - H_{xz} - H_{yz} + H_{xyz} \qquad (2)$$

Depending on how the different variations disturb and condition one another, the outcome of this measure can be either positive or negative. In other words, the interactions among three sources of variance may *reduce* the uncertainty which prevails at the systems level.

For example, the relation between two parents can reduce uncertainty for a child when one parent's answer to a question makes it possible to predict the answer of the other. The parents in this case structure the situation as a family system, but beyond the control of the child at the receiving end. Analogously, processes of codification in terms of cited references and title words used in a document set can reduce uncertainty for future scholars about how to position knowledge claims in new submissions. If this process becomes self-reinforcing, a specific code of communication can be expected increasingly to emerge. Such a code would operate as a latent feedback on the bottom-up construction of new knowledge claims which continue to feed this process substantively.

In other words, the interaction among three variations can endogenously generate a feedback mechanism that reduces the uncertainty within a system. This feedback operates as a recursive loop against the arrow of time (Maturana, 2000). We use this indicator below for the measurement of intellectual organization in document sets. First, we focus on contrasting examples of research in fullerenes *versus* nanotubes as a test case. Thereafter, we extend our reasoning to examples in science studies and the information sciences and focus on citation analysis.

**3. Data**

The indicator is first applied to two sets of documents retrieved from the ISI Web of Science with (a) "fullerene*" and (b) "nanotube*" among their title words. We considered 8,415 documents containing the word fullerene and 17,984 containing the word nanotube. These records were downloaded in the first two weeks of March 2007.

In the case of fullerenes, the analyzed documents cover 20 years, from 1987 to 2006.[4] Of the 8,286 title words in this set, 766 occurred ten or more times, and 3,756 (of the 75,683) cited references were used by more than nine papers. For the case of nanotubes, 15 matrices—corresponding to the publication years 1992-2006—were analyzed containing 976 words, and 6,234 references occurred at least 10 times during the entire period. The configurational information is calculated for each of these matrices in relation to the respective matrix of the preceding year.[5]

---

[4] In 1987, no documents contained the word "fullerene(s)" among their title words.

In a later section, we shall generalize the proposed method by using document sets from our own field of studies—scientometrics—in order to facilitate the interpretation. Is the method so general that it would function even in very differently organized sciences? In order to test this possible generalization, 10,472 documents were downloaded with the word "citation*" among the title words, and 14,805 documents with the word "paradigm*" in March 2008.

The documents using "paradigm" among their title words cover the period from 1956 to 2007. Kuhn's seminal book for science studies entitled *The Structure of Scientific Revolutions* appeared in 1962. Before this date, the word "paradigm" was mostly used in the titles of publications in psychology. Although documents with the word "citation" in the title have appeared since 1922, we did not include titles published before 1964, since the number of documents per year with this title word was erratic during the early period.[6] From 1964 onwards, however, a steady flow of publications with "citation" as title word corresponds to increasing research both in library and information science and in science and technology studies. Can the emergence of research fronts about "paradigm change" or "citation analysis" be distinguished in science studies, the information sciences, and/or their overlap in scientometrics?

---

[5] The routine to compute the configurational information in a set of matrices is available at http://home.medewerker.uva.nl/d.p.lucioarias or http://www.leydesdorff.net/software/synergy/index.htm .

[6] From 1922 to 1964 a total of 83 documents were published using "citation*" as a title word.

Since we will find below that this question can be answered positively for "citation analysis" but not for the concept "paradigm," we further analyzed documents published since 1945 in four relevant journals: (1) *American Documentation*, which in 1969 was renamed the *Journal of the American Society for Information Science* (*JASIS*) and in 2001 became the *Journal of the American Society for Information Science and Technology (JASIST)*, (2) the *Journal of Documentation* (since 1966)*,* (3) the journal *Information Processing & Management* (since 1975) and (4) *Scientometrics* (since 1978). A total of 13,554 documents during the period 1945-2007 are included in this set.[7] The set contains 7,783 references cited by at least four documents, and 941 words which occurred in at least ten titles.[8]

| Set definition | Nr of documents | Nr of title words included in the analysis | Nr of Cited References included | Years | Discussed in section |
|---|---|---|---|---|---|
| "fullerene*" | 8,416 | 986 | 3,756 | 1987-2006 | 4 |
| "nanotube*" | 18,004 | 995 | 5,713 | 1991-2006 | 4 |
| "paradigm*" | 14,805 | 1,011 | 40,487 | 1956-2007 | 5 |
| "citation*" [9] | 10,472 | 965 | 14,989 | 1964-2007 | 5 |
| *Scientometrics* | 2,465 | 793 | 4,502 | 1978-2007 | 6 |
| *JASIST* | 5,034 | 1,008 | 4,422 | 1956-2007 | 6 |
| *IP&M* | 2,541 | 917 | 4,313 | 1975-2007 | 6 |
| *J Documentation* | 3,514 | 792 | 2,757 | 1945-2007 | 6 |

**Table 1:** Descriptive information about the various data sets used in the analysis

---

[7] Of these documents 617 overlap with the set of documents containing "citation*" as a title word.
[8] Our current software has a maximum capacity of 1024 column variables. We can extend beyond this limit, but this would require a substantial investment in software development.
[9] The 83 documents using "citation*" as a title word published between 1922 and 1964 were not included in the analysis.

Table 1 summarizes the various datasets. All title words used in the various analyses were stemmed using the Porter algorithm (Willett, 2006).

**4. Fullerenes and Nanotubes**

The discovery of the new carbon molecules—fullerenes—in 1985 revived an interest in self-assembled nanostructures that preceded the "Nanotechnology Revolution" (Tománek, 2008). In 1990, the invention of the Krätschmer-Huffman generator opened the field to research by providing a method for producing large quantities of purified fullerenes (Baggott, 1994). Sumio Iijima used the generator to replicate the finding of novel carbon tubular architectures that he had discovered ten years earlier with a transmission electron microscope (Iijima, 1991; Harris, 2009, at p. 3). This discovery had the important effect of fullerenizing research in carbon nanotubes (Colbert & Smalley, 2002). Iijima's seminal work allowed pure carbon polymers—first observed by Robert Bacon in 1960—to be understood in the context of the recently discovered fullerenes.

The results for the indicator $\mu^*$ shown in Figure 3 indicate that for "fullerenes" (♦) the prevailing uncertainty measured in bits of information decreased in the early years (1988-1994), but turned positive after 1997, that is, one year after the Nobel Prize for Chemistry was awarded for this discovery. In the case of nanotubes (▲), the lighter line in Figure 3 shows a continuous decline. In this case, $\mu^*$ has been negative since 1994, when single-walled carbon nanotubes—reported for the first time in 1993—became the dominant topic of attention (Iijima, 2002). The ongoing reduction of uncertainty indicates that the emerging feedback arrow of intellectual organization, is gaining in strength.

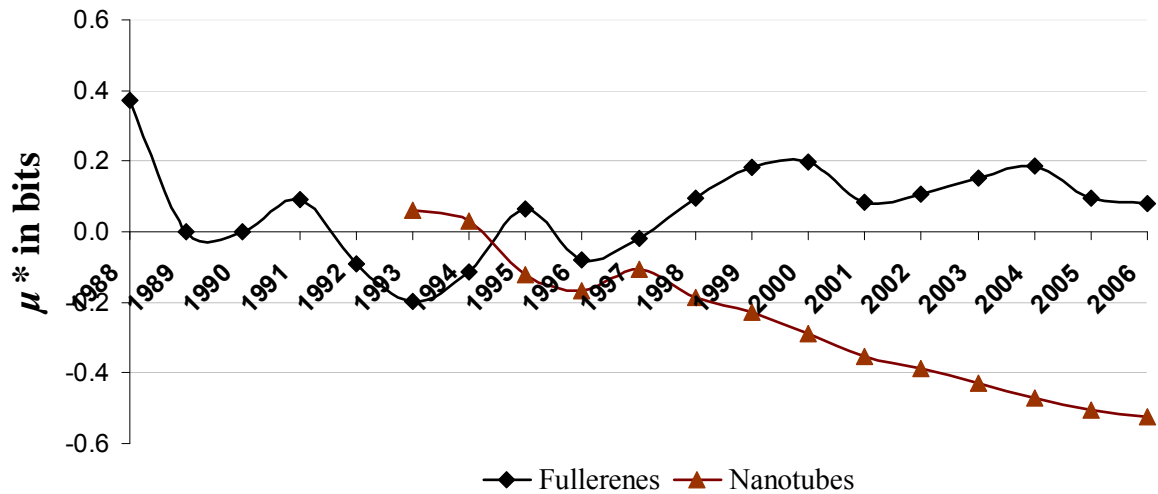

**Figure 3:** Configurational information in bits of information for "fullerenes" and "nanotubes."

After the stabilization of the yearly number of publications with "fullerene*" among their title words (since 1997), the configurational information in this field became positive. Despite the growing number of publications with "nanotube*" in the title and thus a larger variation (Lucio-Arias & Leydesdorff, 2007, at p. 609), the uncertainty in the communication at the systems level is in this case increasingly reduced.

In other words, the indicator suggests that the field of nanotubes is gaining self-organizing momentum in terms of the intellectual organization of the contributions. Even if first observed as a spin-off effect of research in fullerenes, the unique properties of the nanotubes and their promises for potential applications have attracted a great deal of attention (Iijima, 2002; Tománek, 2008). As nanotechnology pioneers transferred their interest increasingly from fullerenes to nanotubes, this meant a shift of attention away from the field of "fullerenes" towards the more recently emerging research program of "nanotubes."

## 5. "Citations" and "Paradigms"

Unlike the natural sciences, in the social sciences theoretical concepts emerge in the literature not in response to discoveries, but in terms of new metaphors (Hesse, 1980). Would the same dynamic of textual organization by intellectual organization hold for this literature, or might it be specific to research fronts in the natural sciences? In order to answer this question we focus on two concepts which have had a central impact on our own field of studies, that is, scientometrics.

Scientometrics can be defined as the quantitative study of scientific communication, and can thus be placed at the crossroads between science studies and the information sciences. Science studies emerged as a specialty in the 1970s after the breakthrough of Thomas Kuhn's (1962) *The Structure of Scientific Revolutions*. Kuhn used the concept of paradigms to describe normal science as opposed to revolutionary science, which would be the exceptional case of *paradigm change* (Popper, [1935] 1959). Kuhn's approach triggered the so-called Popper-Kuhn debate in the philosophy of science (Lakatos & Musgrave, 1970), and in the sociology of science led to the claim that the content of science itself can be made the subject of sociological analysis (Barnes & Dolby, 1970; Bloor, 1976). Since that time, the question about the development of the sciences has turned increasingly into an empirical (instead of philosophical) one (e.g., Hackett *et al.*, 2007).

In an early stage of this debate, Masterman (1970) noted that the term "paradigm" was not strictly defined by Kuhn (1962) and that many meanings of this term remained possible. One should also note that the word "paradigm" already had a clear meaning in linguistics before the First World War. For example, Ferdinand de Saussure (2006) distinguished "paradigms" from "syntagms." In the meantime, however, the concept of "paradigm" has become part of the common vocabulary among scientists when they reflect on their scientific activities. Yet, it has not become a leading concept of a specific research field. In the sociology of scientific knowledge, for example, one distinguishes among different contexts (e.g., Barnes & Edge, 1982) or epistemes (e.g., Knorr-Cetina, 1999). We therefore chose "paradigm" in addition to "citation," since citation analysis has become also a very common notion among scientists, and at the same time functions as a constitutive term for the specialty of scientometrics.

Citation analysis was introduced into the context of the *Science Citation Index* mainly by Eugene Garfield (e.g., 1955). However, citation analysis became a more established routine only after the creation of an experimental version of the *Science Citation Index* in 1962 (Price, 1962, 1965; Garfield, 1972 and 1979; Elkana *et al.*, 1978). The establishment of the journal *Scientometrics* in 1978 marked the beginning of this research area which initially developed in close proximity to other journals in science and technology studies. During the 1990s, the link with journals in the information sciences such as the *Journal of the American Society for Information Science and Technology (JASIST)* became increasingly important for the further development of this specialty (Leydesdorff & Van den Besselaar, 1997; Van den Besselaar, 2000; Leydesdorff, 2007a).

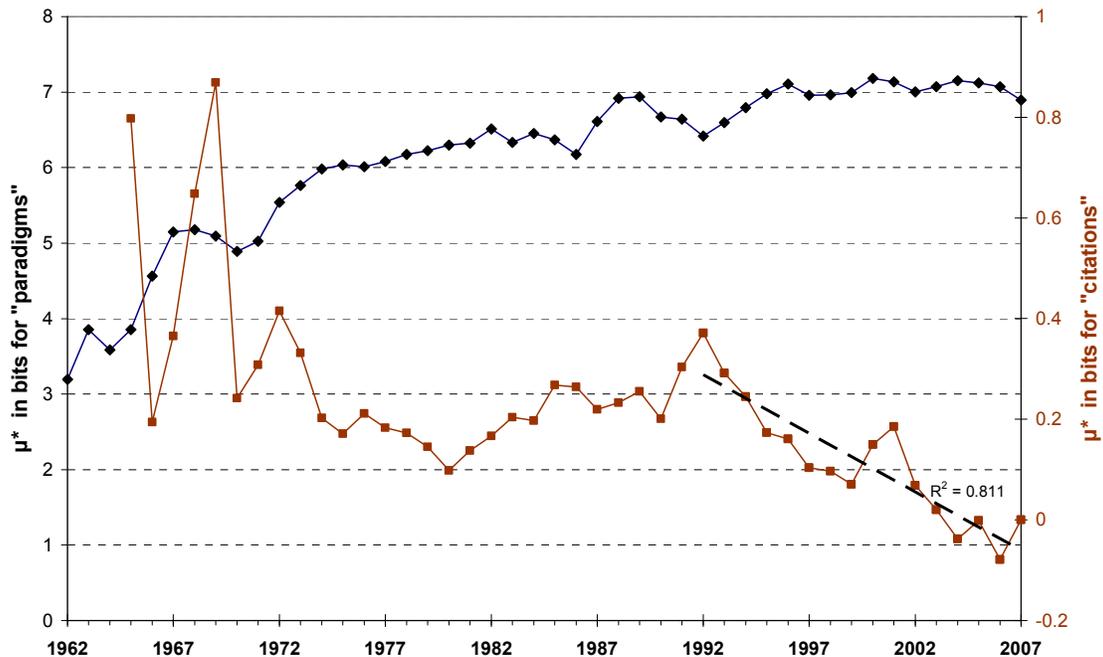

**Figure 4:** Configurational information in bits for "citations" (■) and "paradigms" (♦) as title words.

In Figure 4, the black line (♦) shows the development of $\mu^*$ for documents with "paradigm*" as one of their title words. In this case, the configurational information is always positive and increases over time. According to our hypothesis, this would indicate a lack of intellectual organization between the dimensions of cited references and title words at the aggregated level of the set, even though the yearly numbers of publications increased from 18 in 1962 to 851 in 2007.

The brown line (■) indicates the development of citation analysis. After an irregular pattern in the early years (1962-1975), the introduction of the *Science Citation Index* seems to stabilize this research field during the second half of the 1970s and the 1980s. The 1990s, however, witnessed the emergence of citation analysis as an endogenously organized field of studies. Although continuously declining, the configurational information is still positive as of 2007. In summary, the set with "citation*" among the title words is indicated as increasingly organized. Given this initial result, we decided to generate a more refined representation of the discourse in scientometrics by focusing on all documents published in four relevant journals in the information sciences.

**6. Scientometrics as a further specialization of the information sciences**

We retrieved all documents from the Web of Science which have appeared in the *Journal of Documentation* (since 1966), *American Documentation* (1955), *JASIS* (1969), *JASIST* (2001),[10] *Information Processing & Management* (1975), and *Scientometrics* (1978) as another representation of research in the relevant field of science (see Table 1 above). These journals were selected based on results that indicate scientometrics as a subfield of information science (Van den Besselaar & Heimeriks, 2006).

---

[10] The journal entitled *American Documentation* changed its name to the *Journal of the American Society for Information Science* in 1969, and to the *Journal of the American Society of Information Science and Technology* in 2001.

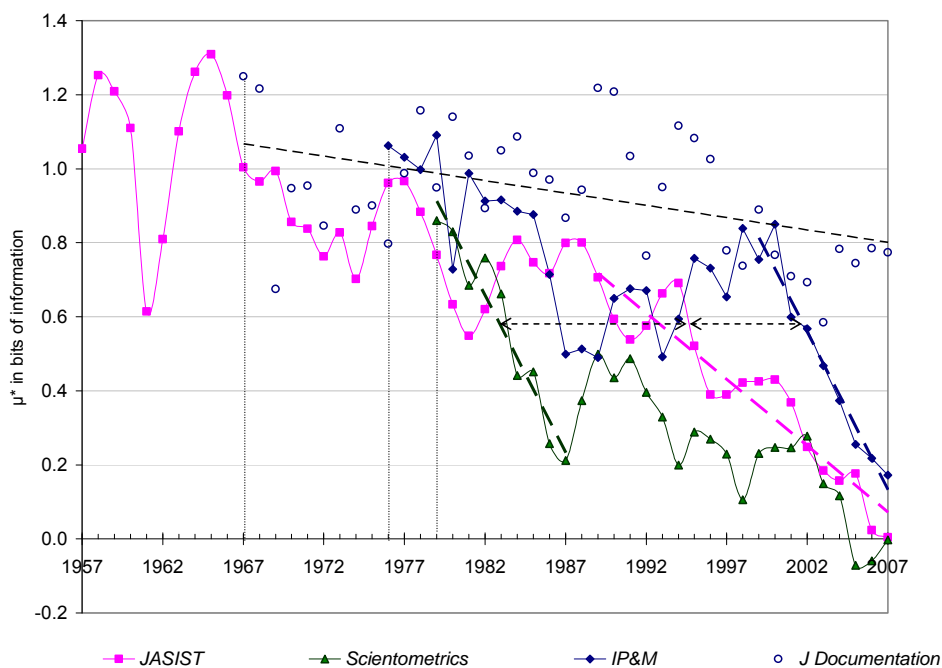

**Figure 5:** Configurational information in bits for journals related to the scientometric discourse.

In Figure 5, the ongoing reduction of uncertainty follows a similar trend for *JASIST* (■), Scientometrics (▲) and *IP&M* (♦) since the late 1970s. Particularly, the publication of *Scientometrics* in 1978 induced a first process of codification at the field level.[11] *JASIST* (■) followed this trend in the 1990s, that is, a decade later, and *IP&M* joins this development in this millennium. The results for the *Journal of Documentation* (○) are added because they indicate that this journal includes research in topics broader than the other three journals, and therefore is less affected by the process of codification and the consequential decrease of uncertainty at the systems level. For this reason, this journal was not included in the estimation of the reduction of uncertainty at the level of the emerging specialty provided in Figure 6.

---

[11] Our method requires at the minimum two years of data, and therefore the first point for *Scientometrics* is indicated in Figure 5 at the date of 1979.

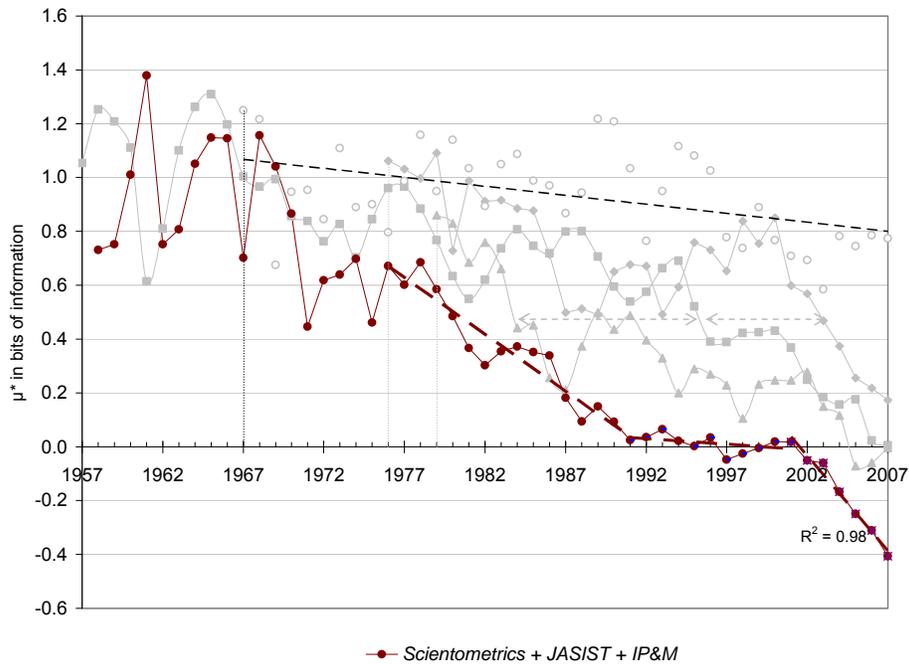

**Figure 6:** Configurational information in bits for the aggregated journals related to the scientometric discourse.

The combined set of *Scientometrics, JASIST,* and *IP&M* (in Figure 6) shows more decline than each of the participating curves (Figure 5). Three periods are suggested which one can recognize as part of the history of the field. First, citation analysis was shaped during the 1980s with a focus in the journal *Scientometrics*. During the 1990s, *JASIST* played as much a central role in codifying this field as *Scientometrics.* The field was relatively broadened and reorganized in the framework of information science and technology. This is also a period in which other themes such as Internet research were topical. Since 2000, citation analysis became an essential tool in the ranking exercises and the evaluation of scientific research. *IP&M* became a third journal with this specific focus.

The curve for "the three journals combined" is steeper than the curves for *Scientometrics, IP&M* or *JASIST* as individual journals. Thus, an additional effect of the interaction at the specialty level is indicated. Indeed, one would expect the discourse (about citation analysis) to develop its code of communication at a level above that of individual journals (Luhmann, 1990; Leydesdorff, 2007b). The emerging code can then be reflected in each of the journals from a specific perspective, although it remains itself a latent variable. One can consider this latent variable as a strategic vector (in this case, intellectual organization) which may develop a feedback to the domain in which it emerges.

Note that the processes under study require a perspective of decades and not a few years. The values of the configurational information even become negative for the combined set after 2002 indicating an increasing synergy. *Scientometrics* is the only journal that already participates in this synergy to the extent of an overall reduction of uncertainty (that is, a negative value of $\mu^*$) since 2006. *JASIST* and *IP&M*, however, can be expected to follow suit in the years to come.

**7. Conclusions and Discussion**

Can the development of discursive knowledge be indicated by using sets of scientific documents? In Lucio-Arias & Leydesdorff (2008), we used Kullback & Leibler's (1951) relative entropy measure to indicate evolutionary turning points in a network of documents, and specified how documents can become "obliterated by incorporation" when their content is increasingly codified in scientific discourse (Merton, 1979). That study, however, used individual documents as units of analysis. Documents—as events—may indicate crucial turning points and main paths to varying degrees.

In this study, we assumed a systems perspective. Science as a complex system can be expected to exhibit non-linear dynamics, including stabilizations along trajectories, and self-organization of socio-cognitive regimes. The latter can be expected to feed back on the historical developments as organizing principles. Our claim is that the information generated from the configuration of title words and cited references in consecutive years provides an indicator of this process of socio-cognitive self-organization.

We first found that after an initial period the research program focusing on fullerenes no longer exhibited the property of reducing uncertainty at the level of the set. In the case of nanotubes, however, since 1993 the reduction of uncertainty has been continuously reinforced. This reduction of uncertainty indicates that intellectual organization of the information is operating as a latent, but increasingly controlling variable.

During this co-evolution between historical stabilization at the trajectory level and evolutionary globalization as a feedback from the regime level, the dependency relations can be expected to change. While the latent variable is first constructed "bottom-up" as a code in the communication, socio-cognitive control in the increasingly codified system is exerted "top-down". The specific codification at the level of the specialty emerges with history, but operates as an evolutionary (selection) mechanism on the historical development which continues to provide the variation. However, the emergent variable can only be measured in terms of its footprints in the observable variation. The *hypothesized* selection mechanism of intellectual organization at the supra-individual level can thus be indicated.

There are no *a priori* reasons why this reasoning would hold only in the natural sciences, although the latter are codified more than (and differently from) the social sciences (e.g., Price, 1970; Bensman, 2008). One can expect codification to be a function of scholarly discourse which is focused on the refinement and elaboration of arguments. However, the social sciences are differently organized; for example, one cannot expect single words to pinpoint discoveries that gave constitutive rise to research specialties as in the case of fullerenes and nanotubes. Innovations are brought about by important new metaphors like the concept of "paradigm" introduced by Kuhn (1962) or the introduction of new instrumentalities like "citation analysis" (Garfield, 1972; Price, 1984).

The concept "paradigm" could not be shown above to indicate a document set displaying increased intellectual organization. On the contrary, the term became diffused among a variety of specialties and has become more common among authors in the understanding and reflexive description of scientific developments. "Citation," however, after an initial period, could be used to indicate decreasing uncertainty, but only since 1992 (Figure 4). Further analysis of the relevant journals in terms of the development of title words and cited references shows decreasing uncertainty at the level of the set of journals since the late 1960s. This ongoing process of codification is visible in the set of documents from *Scientometrics* since the initial year of this journal's publication in 1978.

The document set from *Scientometrics* initiated a more pronounced decrease in uncertainty at the field level during the 1980s, and this codification was reinforced during the 1990s by the contributions in *JASIST* and intensified by *Information Processing & Management* during the 2000s. The three journals combined play an important role in the ongoing development of a new code in communications at the level of the combined sets. The emerging code manifests itself increasingly in terms of specific combinations of title words and cited references which are considered appropriate for contributions containing new knowledge claims. The *Journal of Documentation*, with its focus on library more than information sciences, participated in this development to a lesser extent.

In summary, we model the development of the sciences as an evolving communication of expectations in scientific discourses. The scientific expectations are materialized in texts which report about experiments or other forms of observation. Insofar as the arguments in these texts are increasingly validated in terms of other—previously codified and therefore theoretically relevant—expectations, a research front can be expected to develop. The evolving horizon of expectations can then couple on the textual circulation so that uncertainty on either side of this co-evolution is increasingly reduced because of further specification. The synergy generated by this relative closure can be indicated by the proposed measure.

# References

Abramson, N. (1963). *Information Theory and Coding*. New York, etc.: McGraw-Hill.

Bacon, R. (1960). Growth, Structure, and Properties of Graphite Whiskers. *Journal of Applied Physics,* 31 (2), 283-290.

Baggott, J. E. (1994) *Perfect Symmetry: The Accidental Discovery of Buckminsterfullerene*, Oxford: Oxford University Press.

Barnes, S. B., & Dolby, R. G. A. (1970). The Scientific Ethos: A deviant viewpoint. *European Journal of Sociology,* 11, 3-25.

Barnes, B., & Edge, D. (Eds.). (1982). *Science in Context*. Cambridge, MA: MIT Press.

Bensman, S. J. (2008). Distributional Differences of the Impact Factor in the Sciences Versus the Social Sciences: An Analysis of the Probabilistic Structure of the 2005 Journal Citation Reports. *Journal of the American Society for Information Science and Technology,* 59(9), 1366-1382.

Bloor, D. (1976). *Knowledge and Social Imagery* London, etc.: Routledge & Kegan Paul).

Braam, R. R., Moed, H. F., & van Raan, A. F. J. (1991). Mapping of science by combined co-citation and word analysis. I. Structural aspects. *Journal of the American Society for Information Science,* 42(4), 233-251.

Callon, M., Law, J., & Rip, A. (Eds.). (1986). *Mapping the Dynamics of Science and Technology*. London: Macmillan.

Colbert, D.T. & Smalley, R.E. (2002). Past, Present and Future of Fullerene Nanotubes: Bucytubes. In E. Osawa (Ed) *Perspectives of Fullerene Nanotechnology*. (3-10) Kluwer Academic Publishers, Great Britain.

Elkana, Y., Lederberg, J., Merton, R. K., Thackray, A., & Zuckerman, H. (1978). *Toward a Metric of Science: The advent of science indicators*. New York, etc.: Wiley.

Garfield, E. (1955). Citation Indexes for Science. *Science,* 122(3159), 108-111.

Garfield, E. (1972). Citation Analysis as a Tool in Journal Evaluation. *Science* 178(Number 4060), 471-479.

Garfield, E. (1975). The "obliteration phenomenon" in science—and the advantage of being obliterated. *Current Contents, December,* 22, 396–398.

Harris, F.P.J. (2009). *Carbon Nanotube Science: Synthesis, Properties and Applications*, Cambridge: Cambridge University Press.

Hackett, E., Amsterdamska, O., Lynch, M., & Wajcman, J. (2007*). New Handbook of Science, Technology, and Society*: Cambridge, MA: MIT Press.

Heimeriks, G., Leydesdorff, L., & Besselaar, P. v. d. (2000). Distributed Scientific Communication in the European Information Society: Some cases of "mode 2" fields of research. *Paper presented at Science & Technology Indicators Conference, Leiden, May 24-27, 2000.*

Hellsten, I., Leydesdorff, L., & Wouters, P. (2006). Multiple Presents: How Search Engines Re-write the Past. *New Media & Society,* 8(6), 901-924.

Hesse, M. (1980). *Revolutions and Reconstructions in the Philosophy of Science*. London: Harvester Press.

Iijima, S. (1991). Helical microtubules of graphitic carbon. *Nature*, 354, 56-58.

Iijima, S. (2002). Carbon nanotubes: past, present and future. *Physica B,* 323, 1-5.

Jakulin, A. (2005). *Machine learning based on attribute interactions*. Unpublished PhD. Thesis, University of Ljubljana. Retrieved from http://stat.columbia.edu/~jakulin/Int/jakulin05phd.pdf on March 28, 2009.

Jakulin, A., & Bratko, I. (2004). Quantifying and Visualizing Attribute Interactions: An Approach Based on Entropy. Retrieved from http://arxiv.org/abs/cs.AI/0308002 on March 28 2009.

Knorr-Cetina, K. D. (1999). *Epistemic Cultures: How the Sciences Make Knowledge*. Cambridge, MA: Harvard University Press.

Kroto, H. W., Heath, J. R., O'Brien, S. C., Curl, R. F., & Smalley, R. E.(1985). C60: Buckminsterfullerene. *Nature,* 318, 162-163.

Kuhn, T. S. (1962). *The Structure of Scientific Revolutions*. Chicago: University of Chicago Press.

Kullback, S., & Leibler, R. A. (1951). On Information and Sufficiency. *The Annals of Mathematical Statistics,* 22(1), 79-86.

Lakatos, I. (1970). Falsification and the Methodology of Scientific Research Programmes. In I. Lakatos & A. Musgrave (Eds.), *Criticism and the Growth of Knowledge* (pp. 91-196.). Cambridge: Cambridge University Press.

Lakatos, I., & Musgrave, A. (Eds.). (1970). *Criticism and the Growth of Knowledge* Cambridge: Cambridge University Press.

Law, J. (1986). The Heterogeneity of Texts. In M. Callon, J. Law & A. Rip (Eds.), *Mapping the Dynamics of Science and Technology* (pp. 67-83). London: Macmillan.

Law, J., & Lodge, P. (1984). *Science for Social Scientists*. London, etc.: Macmillan.

Leydesdorff, L. (1989). Words and Co-Words as Indicators of Intellectual Organization. *Research Policy,* 18(4), 209-223.

Leydesdorff, L. (1991). In Search of Epistemic Networks. *Social Studies of Science,* 21, 75-110.

Leydesdorff, L. (1997). Why Words and Co-Words Cannot Map the Development of the Sciences? *Journal of the American Society for Information Science,* 48(5), 418-427.

Leydesdorff, L. (2007a). Mapping Interdisciplinarity at the Interfaces between the *Science Citation Index* and the *Social Science Citation Index*. *Scientometrics,* 71(3), 391-405.

Leydesdorff, L. (2007b). Scientific Communication and Cognitive Codification: Social Systems Theory and the Sociology of Scientific Knowledge. *European Journal of Social Theory,* 10(3), 375-388.

Leydesdorff, L. (2008). Configurational Information as Potentially Negative Entropy: The Triple Helix Model. *Entropy,* 10(4), 391-410; available at http://www.mdpi.com/1099-4300/10/4/391 .

Leydesdorff, L. (2009). Interaction Information: Linear and Nonlinear Interpretations. *International Journal of General Systems,* forthcoming.

Leydesdorff, L., & Sun, Y. (2009). National and International Dimensions of the Triple Helix in Japan: University-Industry-Government versus International Co-Authorship Relations. *Journal of the American Society for Information Science and Technology* 60(4), 778-788.

Leydesdorff, L., & Van den Besselaar, P. (1997). Scientometrics and Communication Theory: Towards Theoretically Informed Indicators. *Scientometrics,* 38, 155-174.

Lucio-Arias, D., & Leydesdorff, L. (2007). Knowledge emergence in scientific communication: from "fullerenes" to "nanotubes". *Scientometrics,* 70(3), 603-632.

Lucio-Arias, D., & Leydesdorff, L. (2008). Main-path analysis and path-dependent transitions in HistCite™-based historiograms. *Journal of the American Society for Information Science and Technology* 59(12), 1948-1962.

Lucio-Arias, D., & Leydesdorff, L. (2009). The Dynamics of Exchanges and References among Scientific Texts, and the Autopoiesis of Discursive Knowledge. *Journal of Informetrics,* 3(2), forthcoming.

Luhmann, N. (1990). *Die Wissenschaft der Gesellschaft*. Frankfurt a.M.: Suhrkamp.

Masterman, M. (1970). The nature of a paradigm. In I. Lakatos & A. Musgrave (Eds.), *Criticism and the Growth of Knowledge* (pp. 59–89). Cambridge, MA: Cambridge University Press.

Maturana, H. R. (2000). The Nature of the Laws of Nature. *Systems Research and Behavioural Science,* 17, 459-468.

McGill, W. J. (1954). Multivariate information transmission. *Psychometrika,* 19(2), 97-116.

Merton, R. K. (1979). Foreword. In: E. Garfield, *Citation Indexing: Its theory and application in Science, Technology and Humanities* (pp. v-ix). New York: Wiley.

Popper, K. R. ([1935] 1959). *The Logic of Scientific Discovery*. London: Hutchinson.

Price, D. J. de Solla (1963). *Little Science, Big Science*. New York: Columbia University Press.

Price, D. J. de Solla (1965). Networks of scientific papers. *Science,* 149, 510- 515.

Price, D. J. de Solla (1970). Citation Measures of Hard Science, Soft Science, Technology, and Nonscience. In C. E. Nelson & D. K. Pollock (Eds.), *Communication among Scientists and Engineers* (pp. 3-22). Lexington, MA: Heath.

Price, D. J. de Solla (1984). The science/technology relationship, the craft of experimental science, and policy for the improvement of high technology innovation *Research Policy,* 13, 3-20.

Pudovkin, A. I., & Garfield, E. (2002). Algorithmic procedure for finding semantically related journals. *Journal of the American Society for Information Science and Technology,* 53(13), 1113-1119.

Saussure, Ferdinand de (2006). *Writings in General Linguistics*, Oxford: Oxford University Press.

Shannon, C. E. (1948). A Mathematical Theory of Communication. *Bell System Technical Journal,* 27, 379-423 and 623-356.

Theil, H. (1972). *Statistical Decomposition Analysis*. Amsterdam/ London: North-Holland.

Tománek, D. (2008). Preface to Carbon Nanotubes: Advanced Topics in the Synthesis, Structure, Properties and Applications, In: A. Jorio, M.S. Dresselhaus, G. Dresselhaus, (Eds.), *Topics in Applied Physics,* Vol. 111, (pp. xi-xiii). Berlin.

Ulanowicz, R. E. (1997). *Ecology, The Ascendent Perspective*. New York: Columbia University Press.

Van den Besselaar, P. (2000). Communication between Science and Technology Studies journals. *Scientometrics,* 47, 169-193.

Van den Besselaar, P., & Heimeriks, G. (2006). Mapping research topics using word-reference co-occurrences: A method and an exploratory case study. *Scientometrics,* 68(3), 377-393.

Willett, P. (2006) The Porter stemming algorithm: then and now. *Program, electronic library and information systems*, 40 (3), 219-223.

Wouters, P. (1999). *The Citation Culture*. Amsterdam: Unpublished Ph.D. Thesis, University of Amsterdam.

Yeung, R. W. (2008). *Information Theory and Network Coding.* New York, NY: Springer. Retrieved on March 29, 2009 from http://iest2.ie.cuhk.edu.hk/~whyeung/post/main2.pdf.